\documentclass[12pt]{iopart}
 \usepackage[margin=1in]{geometry}
\usepackage{float} 
\usepackage{graphicx}
\usepackage{amssymb}
\usepackage{subfigure}
\usepackage{caption}
\usepackage{mathrsfs}

\def\bea{\begin{eqnarray}}
\def\eea{\end{eqnarray}}
\def\be{\begin{equation}}
\def\ee{\end{equation}}

\def\e{\epsilon}

\def\tt{$\tilde{t}$}

\begin{document}
\title{Origins of Thermalization in Semiclassical Cosmology}
\author{Michael Adjei Osei, Arundhati Dasgupta, \\ Physics and Astronomy, University of Lethbridge,\\ 4401 University Drive, Lethbridge, T1K 3M4,\\Canada}

\begin{abstract}
We discuss the effect of `Rindler like' transformation in superspace on the WKB wavefunction in cosmology. In the transformed frame we find the density matrix of the WKB wavefunction is mixed, and this is the sign of a thermal system. This background independent result does not have the same interpretations of a particle thermalization in accelerated frames. One can however show the super space transformation is similar to a double Rindler transformation of real space-time coordinates in the semi-classical regime. We also discuss the implications of the super-space transformation in loop quantum cosmology. 
\end{abstract}

\section{Introduction}
We are trying to understand the physics of the distant past for the origins of our current day observations. We live surrounded by a universe with stars, galaxies, celestial objects, all moving away from each other, causing the entire universe to expand at an accelerating rate \cite{Peebles:2022akh}. The visible matter is embedded in a stochastic background of gravitational and electromagnetic radiation. The latter is at $2.78 ^\circ$ and is known as the cosmic microwave background. The stochastic gravitational waves are yet to be detected directly, but there is evidence for their existence. If we try to use Einstein's General Theory of Relativity  to explain these observations then the mathematical model, known as $\Lambda$ CDM cosmology, predicts the existence of dark energy and dark matter \cite{Lahav:2020hzh}. There has been immense work in trying to understand the origins of cosmic evolution \cite{ashtekar2015generalized}; however, some of the physics of the early universe is yet to be tested, and we contribute a new perspective discussion to what might be the correct way of obtaining the physics of today. Our work bridges quantum cosmology \cite{kiefer2012emergence, halliwell1991introductory}, thermal effects \cite{Dasgupta:2015ufa} of acceleration as well as loop quantum cosmology\cite{ashtekar2015generalized}. This work establishes a fundamental link between accelerated frames and thermalization of the semiclassical universe, analogous to the Unruh effect \cite{takagi1986vacuum, fulling1973nonuniqueness}. By mapping Rindler transformations to superspace (Section 3), we show the Wheeler-DeWitt wavefunction undergoes Bogoliubov mixing (Section 4), leading to a thermal density matrix. This suggests quantum gravity may encode thermodynamic irreversibility in cosmological evolution. The use of the `accelerated frame' in super-space is to implement thermalization of the quantum wavefunction in a background independent way. The use of transformation of fields in scalar field cosmology can be found in \cite{gielen23}. We use a `Rindler transformation' in super-space, though the interpretation of the transformations as `acceleration' in super-space is not obvious (apart from field redefinitions). To connect with our usual accelerated observer, we use the semi-classical solutions to the scale-factor and the scalar field. In this approximation the transformation in super-space can be interpreted as a double-Rindler-like transformation of the background co-moving frame of the cosmology. This therefore is indeed a symmetry of the Gravitational system in the semi-classical limit.

 Our calculations are theoretical, and the Wheeler-De Witt equation (WDE) formulated for canonical gravity is used to obtain the semi-classical universe. The question we are asking is what is the correct frame for the human observer today in cosmology?  Is it co-moving frame as we have formulated, or our earth is accelerated (in a complicated way) wrt the `real' co-moving observer of our cosmic evolution formulation.  Our eventual aim is to connect with observational data, including finding origins of dark energy and dark matter as future work.

 In the next section we describe the WDE for scalar field cosmology. We use this for convenience, but our formulation can be extended to other cosmologies. In the third section we discuss the Rindler-like transformation in super-space. The fourth section is a discussion of the thermalization using the density matrix built using the WDE semi-classical wavefunction. As there is no particle interpretation the usual Planck distribution in accelerated frames as derived in \cite{Unruh:1976db} and \cite{hawking1975} is not used. However, our derivation is relevant for quantum fields for accelerated observers in Minkowski background as well \cite{Dasgupta:2015ufa}.  We conclude in the fifth section, with a discussion about the connection of this work with loop quantum cosmology.
\section{Theoretical Framework}

\subsection{Superspace and WKB Approximation}
If we have the cosmological metric in conformal time of an isotropic, homogeneous, asymptotically flat space-time, it takes the form \cite{cosmo1}
\be
ds^2 = a^2(\tau)(-d\tau^2 + dx^a dx_a).
\ee

We add matter to this system as a scalar field $\phi$, to facilitate a matter-driven cosmology. The Hamiltonian takes the form:

\be
H_{\rm RW}= -\frac{p_a^2}{24 a} + \frac{p_{\phi}^2}{2 a^3} +a^3 V(\phi).
\ee
$p_{\phi}$ and $p_a$ are the momentum corresponding to the scalar field and the scale factor. The details of the derivation can be found in \cite{Isham:1992ms}, \cite{Vilenkin:1987kf}. $V(\phi)$ is a potential for the scalar field and in the first approximation, we set the potential to zero, and analyze a WKB solution \cite{Vilenkin:1987kf}. 

The WDE in presence of a scalar field is obtained, the equation takes the form (\cite{Isham:1992ms}) 
\be
\hat{H}_{\rm RW} \psi (a ,\phi) = \left(-\frac{\hat{p}_a^2}{24 a} + \frac{\hat{p}_{\phi}^2}{2 a^3}\right) \psi(a,\phi)=0.
\ee

Before we lift the above to the quantum operator, one gets operator ordering ambiguities. So, we order the WDE equation in the following form
\be
\left(-\frac1{12} \hat{a} \hat{p}_{a} \hat{a} \hat{p}_a + \hat{p}^2_{\phi}\right)\psi(a,\phi)=0.
\ee
Using the operator representation $p_{a}= i \frac{\delta}{\delta a}$  and $p_{\phi}=i \frac{\delta}{\delta \phi}$, one gets 
\be
\left(\frac{a}{12} \frac{\delta}{\delta a}\left(a \frac{\delta \psi}{\delta a}\right) - \frac{\delta^2 \psi}{\delta \phi^2}\right)=0.
\ee
Using the redefinition of the coordinate in superspace (space of $a,\phi$) as $ \xi= \ln(a)$, one gets the WDE as
\be
\left(\frac1{12} \frac{\delta^2 \psi}{\delta \xi^2} - \frac{\delta^2 \psi}{\delta \phi^2}\right)=0.
\ee
One can absorb the constant 12 into a redefinition of $\xi$ and eventually obtain the WDE on superspace as
\be
\frac{\delta^2 \psi}{\delta \xi'^2} - \frac{\delta^2 \psi}{\delta \phi^2}=0.
\label{eqn:wde}
\ee

Using the WKB approximation of Vilenkin \cite{Vilenkin:1987kf}, a semi-classical solution is of the form
\be
\psi(\xi',\phi)= e^{i k(\xi'\pm\phi)}.
\ee

One can draw this wave function in the $-\infty<\xi'<\infty$ and $-\infty<\phi<\infty$ `causal diamond' with null lines extending from null past to null infinity.
\begin{figure}
    \centering
    \includegraphics[width=0.5\linewidth]{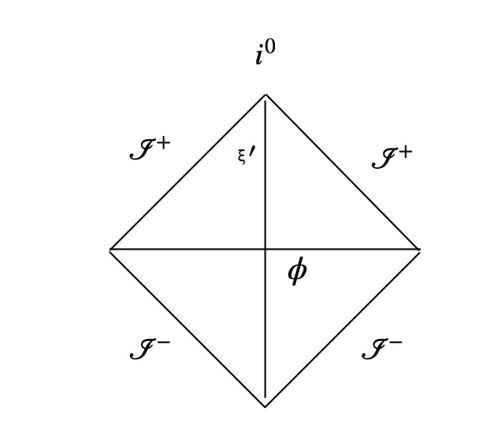}
    \caption{This diagram represents the causal diamond in the DeWitt superspace, illustrating the coordinates $\xi '$ and $\phi$. The null lines divide the space into four regions: $\mathscr{I}^+$) (future null infinity), $\mathscr{I}^-$ (past null infinity). The center of the diamond corresponds to the present moment in superspace. This visualization aids in understanding the wave function's behavior with `lightfront' waves.}
    \label{fig:enter-label}
\end{figure} \\

 \begin{figure}[H]
  \centering
\includegraphics[scale=0.65]{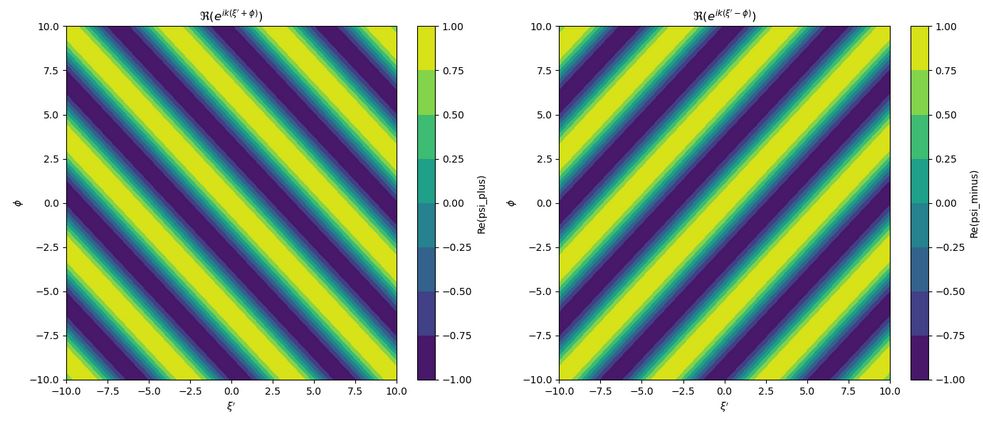}
\caption{This diagram shows the behavior of the wave function ($ \psi(\xi', \phi) =  \e^{ik(\xi' \pm \phi)} $) in the $ \xi' $ and $ \phi $ frame. The plot illustrates the wave function's amplitude and phase across the $\xi'$ and $\phi $ coordinates. The wavefunctions are similar to massless waves in the superspace, with lightfronts at 45 degrees.}
\end{figure}.\\

 \begin{figure}[H]
  \centering
\includegraphics[scale=0.65]{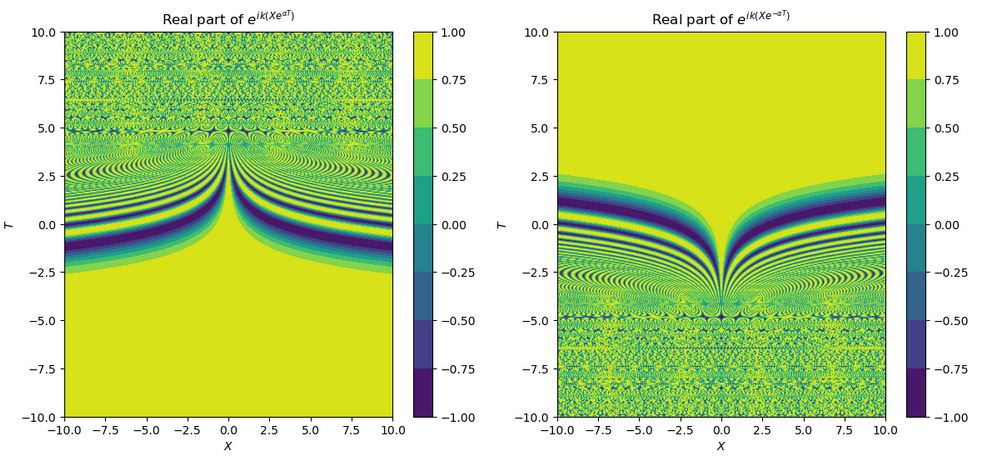}
\caption{This diagram shows the behavior of the wave function ($ \psi(X, T) =  \e^{ik(X e^{\pm T})} $) in the $ X $ and $ T $ frame. The plot illustrates the wave function's amplitude in the $X$ and $T $ coordinates, depicting the intricate patterns formed by the interaction of these variables. The wavefunctions are considerably different than massless waves in the superspace as in the previous figure.}
\end{figure}.\\

From the D'Alembertian, one can identify the ($\xi',\phi$) coordinates as that of a super-space with a flat metric. As one can transform these coordinates, as in real flat space-time, one can obtain the wave-function in different reference frames. In real flat space-time, these could be boosts which keep the Lagrangian invariant. As the theory of the wavefunction in super-space is not formulated as diffeomorphism invariant, these transformations are expected to change the wave function. However, in the semi-classical limit the superspace transformations are induced from a double Rindler coordinate transformation in real space-time, which keep the underlying Hamiltonian invariant. We discuss this in the next section.

 \section{Transformation in Super-Space and Rindler Coordinates}
We can make Rindler transformations in super-space, to `accelerating coordinates' ($X,T$).
\be
\xi'= X \sinh(\alpha T);  \ \ \phi= X \cosh(\alpha T).
\ee

\begin{figure}
\centering
\includegraphics[scale=0.6]{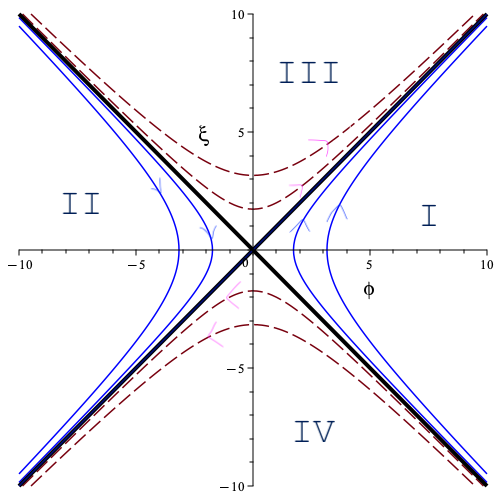}
\caption{The Minkowski Diamond and the Rindler Horizons at $\xi'=\pm \phi$. The Regions I and II have the same coordinate transformations to (X,T). For regions III and IV one implements transformation to another set of coordinates $\tilde X, \tilde T$ which are analytic continuation of ($X,T$)}
\label{fig:rindler}
\end{figure}

Here we have used the same notation as coordinate transformation in real space-time though they are transforming `fields' $a,\phi$. In the $X, T$ frame only wedges of the super-space causal diamonds will be accessible controlled by the parameter $\alpha$. 
As is obvious from the above transformation, $\phi> \xi'$ for the coordinate transformation to exist. As shown in the Figure (\ref{fig:rindler}), the transformation is valid only in regions I, II. Further the I and II regions, are entire universes for the X, T coordinates, with time flowing forward in I, and backward in II. In Region I, time $T$ is given as 
\be 
T= \frac1{2 \alpha} \ln\left(\frac{\phi+\xi}{\phi-\xi}\right)
\ee
$T\rightarrow \infty$ as $\xi=\phi$ and $T\rightarrow -\infty$ as $\phi\rightarrow \infty$, and therefore the clock in the accelerated observer frame never crosses the $\phi=\xi$ surface.

For region III and IV, the coordinate transformations have to be different, lets say $(\tilde{X}, \tilde{T})$, and they would be such that
\be
\xi'= \tilde{X} \cosh(\alpha \tilde{T}); \ \ \  \ \ \ \phi= \tilde{X} \sinh(\alpha \tilde{T}).
\ee

Note that the $\tilde{X}=c$ curves are now space-like, and one can use analytic continuation such that
\be 
\tilde{X}= i X ; \ \ \ \ \ \ \ \tilde{T}= T - i \frac{\pi}{2 \alpha}.
\ee

The WDE equation for the new coordinates is
\be
-\alpha^2 X \frac{d}{dX} \left( X \frac{d \Psi}{d X}\right) + \frac{d^2 \Psi}{dT^2}=0.
\ee
The solution to this equation is : \be \Psi(X,T) = \bigg(e^{i k' \left(\pm \frac{1}{\alpha} \ln(X) + T\right)}\bigg). \ee
If we implement the coordinate transformation of the super-space on the wavefunction obtained in the $\xi',\phi$ coordinate, one has
\be
\psi(X, T)= e^{i k(X e^{\pm \alpha T})}.
\ee

Making a transformation of $ u (v) \ = \ \frac{1}{\alpha} \ln(X) - (+) T$ the wavefunction for the static frame and the accelerated frame could be expressed as:
\be
\Psi(u) = e^{i k' u (v)},
\ee

\be
\Psi(u) = e^{i k e^{\alpha u(v)}}.
\ee

In the above the nature of the wavefronts (left moving or right moving) determines whether one uses the u or the v coordinate. In the subsequent discussions we use the $u$ coordinate, but it is to be noted that analytically continuing the coordinates to other regions can switch the nature of the wavefunction dependence. Our subsequent discussions are therefore strictly restricted to Region I.

 \begin{figure}[H]
  \centering
\includegraphics[scale=0.7]{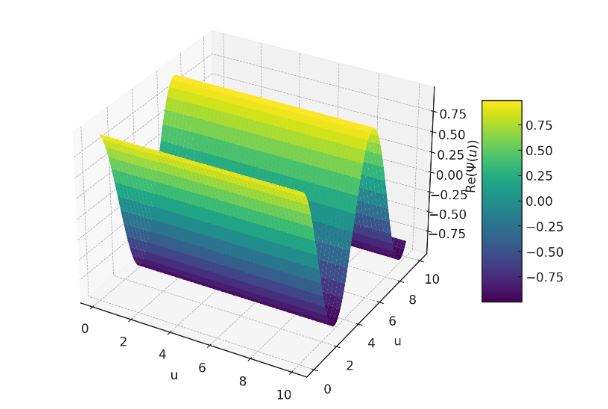}
\caption{This plot visualizes the oscillatory behavior of the wavefunction for the static frame over the  range of $ 0 < u < 10$.}
\end{figure}.

 \begin{figure}[H]
  \centering
\includegraphics[scale=0.7]{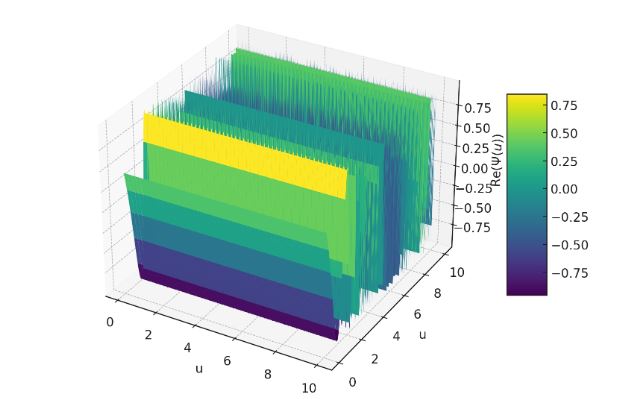}
\caption{This plot illustrates the physical behavior of the wavefunction in the accelerated frame within the Right Rindler Wedge, demonstrating the exponential oscillatory nature of the wavefunction over the range of $ 0 < u < 10$.}
\end{figure}.

\subsection{Double Rindler Interpretation and Cosmological Implications}

If we now interpret the transformation in super-space as due to the acceleration of the observer in real space-time, the quantum vacuum of the universe will have entropy creation as predicted in \cite{Dasgupta:2015ufa, prigogine1980being}. Is this interpretation plausible? If we are in the semi-classical regime, the classical solutions of the scale factor and the scalar field are functions of the space-time coordinate. 
If we see the scale factor for the scalar energy-momentum tensor as a function of co-moving coordinates, then \cite{Vilenkin:1987kf},
\be
a(t)\propto (t-t_0)^{1/3},
\label{eqn:class1}
\ee
with the scalar field as
\be\phi(t) \propto \frac{1}{3} \ln (t-t_0). \label{eqn:class2}\ee
If one makes a transformation to Rindler-like coordinates $(T, X)$ which represent `accelerated observers' in  the cosmological superspace, for the conformal coordinates $\xi=\ln a(t)$, then the coordinates defined earlier, transform as:

\be
\xi'\approx\frac13 \ln(t-t_0) \rightarrow T,X \ \ \phi\approx \frac13 \ln(t-t_0)\rightarrow T, X
\ee
As defined previously ($\alpha$ is a constant)
\be
\xi'= X\sinh(\alpha\  T) \  \ \phi= X \cosh(\alpha\ T).
\label{eqn:rindler}
\ee

Note that the coordinates $\ln a(t)$ and $\phi$ are coordinates in the super space, and therefore these are not interpretable as coordinate transformations of a `real' space-time metric. However, as we are dealing with semi-classical physics, we can use the classical solutions of these fields in terms of the co-moving time.
If we use the classical solutions as in Equations(\ref{eqn:class1}, \ref{eqn:class2}), we can find the following, in the limit of $\alpha >>1$

\be
t-t_0\approx \exp\left( 3 X \exp(\alpha T)\right).
\label{eqn:rindler1}
\ee

For constant $X$, this is a `double Rindler' transformation. It can also be interpreted as a transformation to a frame where the acceleration is not constant, but is exponential function of the static time $T$. 
This interpretation can be seen using a Lorentz transformation ($t,x\rightarrow t',x'$) to a frame which is moving with a velocity v 
\begin{eqnarray}
    t' &=& \cosh(v) t + \sinh(v) x\\
x' &= & -\sinh(v) t + \cosh(v) x
\end{eqnarray}

At $t \approx 0$, this would be
\begin{eqnarray}
    t' & = & + \sinh(v) x\approx e^{\alpha t} x,\\
x' & = & \cosh(v) x \approx  e^{\alpha t} x.
\end{eqnarray}

If the system has constant acceleration of $\alpha t=v$ and $\alpha >>1$. We then make another transformation which is also a boost to new coordinates (t",x") with a parameter $\beta=3$:
\begin{eqnarray}
t'' &=& \exp(\beta t') x'\approx \exp\left(3x \ e^{\alpha t}\right)x', \\
x'' &=& \exp(\beta t') x' \approx \exp\left(3x\ e^{\alpha t}\right)x'.
\end{eqnarray}
This transformation can be interpreted as a `double Rindler' transformation. Previously such double Rindler coordinate transformations have been implemented in Minkowski background quantum fields \cite{doubler}.
In our example, co-moving time $t$ is related to $T$ by a `double Rindler' like transformation as in Equation \ref{eqn:rindler1}. In this interpretation too the $\alpha$ is the only parameter and it measures the acceleration in the first Rindler transformation.

Note, our wavefunction/field is defined in super-space, and our transformations are at most `field re-defintions' as in \cite{gielen23}. Here, the wavefunction of the universe shows a thermal behavior which we discuss next, and can be found rooted in the `Unruh' effect \cite{Unruh:1976db}. However, in super-space, the scale-factor and the scalar field are both quantized, and the entropy production involves both the matter and the gravitational degree of freedom. As pointed in \cite{prigcosmo}, in a quantized space-time scenario, the scalar field isn't in a fixed background, but is interacting with the quantum geometry.  Infact allowing particle exchange of gravity and matter leads to singularity avoidance \cite{prigcosmo} in the cosmic evolution. Next we try to investigate the thermal nature of the WDE wavefunction. \\
From Figure 2 and Figure 3, one can see that the wave function behaves differently after it was transformed in the $T,X$ frame from the ordinary light cone wavefronts. We now progress to analyze the static wave function mathematically in $X,T$, frame and we do that by employing the Bogoliubov transformation.

\section{Emergence of Thermalization}
\subsection{Bogoliubov Transformation}
 We examine the WDE solutions in the accelerated frames' observer.
If we identify the modes of the wave function in the $T$ coordinate as 
\be
\psi_T= \exp(i k' u).
\label{eqn:rindler}
\ee
a Bogoliubov transformation relates the two wave-functions \cite{Unruh:1976db,crispino2008unruh}.

As we can use the `light-like' coordinates in $\xi',\phi$ space, which we have defined previously, and setting the $V(\phi)=0$ gives the D'Alembertian a null structure, we find the Bogoluibov coefficients, using the `null' modes. 
To find the Bogoliubov coefficient we find the inner product of the mode $\exp(i k u)$ where $u=\frac1{\alpha} \ln X -T$ wrt to the static wave function which has the form $\exp(i k \exp(\alpha u))$.  This therefore allows us to write the static wavefunction as a linear combination of the `accelerated' modes of the Universe. Note that, the interesting aspect is the appearance of the negative and positive frequency mode mixing.  The inner product used is:
\be
\langle \Psi_1, \Psi_2 \rangle = i \int_{\Sigma} d u \left( \Psi_1^* \partial_\mu \Psi_2 - \Psi_2 \partial_\mu \Psi_1^* \right),
\ee
And using the two wavefunctions as in static and accelerated frames:
\be
\alpha_{kk'}=\langle \Psi_1, \Psi_2 \rangle = i \int_{\Sigma} du \left( e^{-ik' u} \cdot ik \alpha e^{\alpha u} e^{ik e^{\alpha u}} - e^{ik e^{\alpha u}} \cdot (-ik' e^{-ik' u}) \right).
\ee

\be
\alpha_{kk'} = \frac{2}{i} \frac{\Gamma(1-i k'/\alpha)}{(-ik)^{-i k'/\alpha}}
\ee
Similarly the inner product of the static wavefunction with the negative frequency Region I mode is obtained by taking $k'\rightarrow -k'$.

\be
\beta_{kk'} = \frac{2}{i} \frac{\Gamma(1+i k'/\alpha)}{(-i k)^{i k'/\alpha}}.
\ee

In case of quantum fields one can verify that the above can provide the origin of the Boltzmann distribution of black body radiation when we calculate the modulus of $\alpha_{k k'}$:

\be
|\alpha_{k k'}| \ \approx \  \exp \left({  \frac{ \pi k'}{ \alpha}} \right) \ |\beta_{k k'}|.
\ee
This can be interpreted as thermalization 
\cite{Unruh:1976db,hawking1975, Birrell:1982ix,birrell1984quantum} in the context of a scalar quantum field.
For the wavefunction of the universe as both coefficients($\alpha_{k k'}, \beta_{k k'}$) are non-zero, it shows a mixing of negative and positive frequencies.

 There are no creation and annihilation operators, but we find a very interesting technical result about the Bogoliubov coefficients.  This leads to our conclusion in the next section about the `mixed' nature of the density matrix built out of the static wavefunction in the accelerated frame.
We observe that the integral over the $\alpha_{kk'}$ is finite, but the integral for $\beta_{kk'}$ is infinite.
\be
\int_{0}^{\infty} dk' |\alpha_{kk'}|^2 = 4\int \frac{|\Gamma(1-ik'/\alpha)|^2}{(e^{i \pi})^{-i k'/\alpha}} dk'= \frac{4 \pi}{\alpha}\int \frac{k' e^{-\pi k'/\alpha}}{\sinh(\pi k'/\alpha)}  d k'=\frac{\pi \alpha}{3}
\ee
and 
\be
\int_{0}^{\infty} dk' |\beta_{kk'}|^2 = 4 \int \frac{|\Gamma(1-ik'/\alpha)|^2}{(e^{i\pi})^{i k'/\alpha}} dk' =\frac{4 \pi}{\alpha}\int \frac{k' e^{\pi k'/\alpha}}{\sinh(\pi k'/\alpha)}  d k'=
\infty
\ee
This result can be regulated using a finite cutoff for the frequency $k'$. If we set that as $\Lambda$, then
\be
\int_0^{\Lambda} dk' |\beta_{kk'}|^2 =\frac{4\Lambda^2 \alpha}{\pi} + \frac{\pi \alpha}{3} + O(\exp(-2 \Lambda))
\ee

This result is related to infinities encountered in quantum field theory, due to a continuum of modes. The infinity can also be attributed to the fact that as $u\rightarrow \infty$, which is where the `Rindler' horizon is located, the Rindler modes pile up with an infinite number of them accumulating into the static mode.

\subsection{Density Matrix Formalism and Mixed States}
Following the formulations of \cite{Dasgupta:2015ufa}, \cite{prigogine1980being} and , we try to find if there is a thermalization process involved. To study, we use the density matrix formalism.
\be
\Lambda=|\psi><\psi|
\ee

  To check for thermalization we need to see if we have a mixed state in the new basis \cite{hu2008stochastic}. As we are aware the state $\psi$ is a linear combination of the basis states of $|k>\sim e^{iku}, |-k>\sim e^{-ik u}$ in Region I. Is the state mixed in the new basis? By definition, the static state density matrix is
\begin{eqnarray}
\rho = |\psi\rangle\langle\psi| &=& \int dk'\, dk'' \big(
\alpha_{k k'} \alpha^*_{k k''} |k'\rangle\langle k''| + \alpha_{k k'} \beta^*_{k k''} |k'\rangle\langle -k''| \nonumber\\
&& \quad + \beta_{k k'} \alpha^*_{k k''} |-k'\rangle\langle k''| + \beta_{k k'} \beta^*_{k k''} |-k'\rangle\langle -k''|
\big)
\label{eqn:density}
\end{eqnarray}

If we take the inner product of this with the states $|m>$, which are such that $<m|n>=\delta(m-n)$, one gets the elements of the density matrix to be
\begin{equation}
    \rho_{mn}= \alpha_{km}\alpha^*_{kn}+\alpha_{km}\beta^*_{k-n} +\beta_{k -m} \alpha^*_{k n} +\beta_{k-m}\beta_{k -n}^*
\end{equation}
As there is a normalization freedom, one can use that to set ${\rm Tr} \rho=1$. Note that one might find it odd that the density matrix defined using the outer product of one state is not pure, when we aren't `explicitly tracing' over a part of the system. However, if one sees that we are using the modes of the accelerated frame, then only a part of the static wavefunction is contributing to the trace of the density matrix. This can also be realized from the plots of the wave-function in the accelerated frame, where the modes of the accelerated frame pile up at the horizon. The coordinates $X$ is defined only within the wedge $\phi>\xi'$ of the original causal diamond (Region I of Figure(\ref{fig:rindler})). Thus the entire domain of definition of $u$ is exhausted within the wedge. The Bogoluibov coefficients are valid only for $\phi>\xi'$ and therefore the above mode expansion of the density matrix in Equation (\ref{eqn:density}) is valid only for the `Rindler' wedge.  
We find ${\rm Tr}(\rho)$ rather carefully.
\be
{\rm Tr} (\rho)= \int_{0}^{\infty} \  dm <m|\rho|m> = \int_0^{\infty} \ dm \ |\alpha_{km}|^2 + \int^{\infty}_0 \ dm \ |\beta_{k -m}|^2 
\ee
This however, is a finite quantity as $\int_{0}^{\infty} dm \  |\beta_{k-m}|^2$ is well defined.
Thus we can use a normalization constant to set the trace to 1. 
If we use a normalization constant $A_0$ one gets
\be
{\rm Tr}(\rho)= \frac{|A|^2\  2 \pi \alpha}{3} \ = 1 \rightarrow |A|^2= \frac{3}{2 \ \pi \alpha} 
\ee

To compute $\rho^2$ we use integrals
\begin{eqnarray}
\rho^2 &=& \int dk' dk''\left[\alpha_{kk'} \left( \int (\alpha_{kl}^* \alpha_{kl} +     \beta_{kl}^* \beta_{kl}) dl \right) \alpha^*_{kk"} |k'><k''| \right.\nonumber\\
&& + \alpha_{kk'} \left( \int (\alpha_{kl}^* \alpha_{kl} + \beta_{kl}^* \beta_{kl}) dl \right) \beta^*_{kk''}|k'><-k''| \nonumber\\
&& + \beta_{kk'} \left(\int (\alpha_{kl}^* \alpha_{kl} + \beta_{kl}^* \beta_{kl}\right)\alpha_{k k''} |-k'><k''| \nonumber\\
&& + \beta_{kk'} \left( \int (\alpha_{kl}^* \alpha_{kl}+ 
 \beta_{kl}^* \beta_{kl}) dl \right) \beta_{kk''}^* |-k'><-k''|\left.\right].
\end{eqnarray}
As the Bogoliubov coefficients are explicitly known, one can calculate the integrals exactly. They are infinite, as expected, due to $l\subset[0,\infty]$. However, integrals can be regularized using ultraviolet cutoff.
Using the fact that:
\be
\int dl \left(|\alpha_{kl}|^2 + |\beta_{kl}|^2 \right)= 1+ \frac{6}{\pi^2}\Lambda^2,
\ee
one finds $\rho^2\neq \rho$, as 
\be
\rho^2= (1+ \frac{6}{\pi^2}\Lambda^2) \rho,
\ee
and
one can show that using the same normalization as ${\rm Tr} \rho$,\  ${\rm Tr} \rho^2 \neq 1$. This discrepancy in a overall factor between the matrix and it's square cannot be normalized away. As $\Lambda \rightarrow \infty$ the factor is infinite, and is a sign that the static density matrix is not complete in the accelerated observer's frame, and is not pure according to the standard definition. This result is re-affirming the observation that quantum states in accelerated frames is a non-unitary transform of the quantum state observed in static frames \cite{Dasgupta:2015ufa, dasgupta2019}. A density matrix whose square is a multiple of it, and is mixed has been studied, e.g. $\rho_{ij}=(1/2)~ \delta_{ij}\rightarrow \rho^2 =(1/2) ~ \rho$. We find a similar system, except the proportionality factor is infinite in the limit $\Lambda \rightarrow \infty$.  

\section{Conclusion and Connection to Loop Quantum Cosmology}
Due to the fact that the above discussion pertains to semiclassical solutions of WDE cosmology, we try to find a parallel of the super-space transformations in Loop Quantum Cosmology (LQC) \cite{barrau2016some}. Through canonical quantization, we derived the Wheeler-Dewitt equation in terms of the LQC variables \cite{bojowald2008loop}. A detailed derivation can be found in the Appendix.

In LQC the basic variables are defined as the connection and the `triad' (momentum)
as 
\begin{equation}
    A_a^i = c \delta^i _a \  \  \  \  E^a_i = p \delta^a_i
\end{equation}
where in terms of the classical metric $c={\dot a}{a^2}$ and $p= a^2$. The fiducial volume factor and the Immirzi parameter have been set to 1 in this analysis.

The WDE in Loop Quantum Cosmology using Fourier analysis of the loop representation is given by: \begin{equation}
\frac{\partial^2 \hat{\psi}(p', \phi)}{\partial \phi^2} +  \frac{1}{2\mu^2} \hat{\psi}(p', \phi) - \frac{1}{4\mu^2} \left( \hat{\psi}(p' - 2\mu, \phi) + \hat{\psi}(p' + 2\mu, \phi) \right) = 0.
\end{equation}

We can show that the Vilenkin wavefunction satisfies the above equation in the discretized limit. The $p'=\ln (p)$ is discretized with a spacing of $\mu$. We can implement the `accelerated frame' transformation in the LQC coordinates, too. This would amount to boosting a discretized coordinate, and there has been considerable discussion on what boosting means in quantum cosmology framework \cite{liv22}. 

 In this paper, we have shown that a particular accelerated frame in cosmology is the same as `coordinate transformation' in the DeWitt superspace. This transformation results in a Bogoliubov transformation of the semi-classical wave function of the universe defined on the DeWitt superspace. The transformation is Rindler like, and is motivated from the form of the D'alembertian for a particular operator ordering of the super-space WDE. Now in super-space, there is no requirement that a coordinate transformation keeps the wave-function invariant. However, what we have implemented the matter fields can transform into gravitational degrees of freedom, and there will be `particle' and entropy exchange between the fields. The particular transformation we have implemented thermalizes the total Universe and this quantum state must be evolving today. The existence of matter-gravity exchange and the cosmic evolution has to be investigated further, and real physical implications of singularity resolution and emergence of large-scale structure of today is in progress.

\section*{Acknowledgement}
We thank Prof. Faraoni, Prof. Vos for useful discussions, and Anusha Vasaikar for collaboration.\\

\appendix

\section*{Appendix A. WKB Solution to the Superspace D'Alembertian}

We consider the Wheeler-DeWitt equation in superspace:
\begin{equation*}
\frac{\partial^2 \Psi}{\partial \xi'^2} - \frac{\partial^2 \Psi}{\partial \phi^2} = 0 \  \qquad (A.1). 
\end{equation*}

We seek a WKB-type solution of the form:
\begin{equation*}
\Psi(\xi',\phi) = A(\xi',\phi) e^{iS(\xi',\phi)}, \ \qquad(A.2)
\end{equation*}
where $A(\xi',\phi)$ is a slowly varying amplitude and $S(\xi',\phi)$ is a rapidly varying phase.

Substituting (A.2) into (A.1) and expanding derivatives, and keeping only leading-order terms in the WKB approximation, we find:
\begin{equation*}
\left( \frac{\partial S}{\partial \xi'} \right)^2 - \left( \frac{\partial S}{\partial \phi} \right)^2 = 0. \
\qquad{A.3}
\end{equation*}
Thus,
\begin{equation*}
\frac{\partial S}{\partial \xi'} = \pm \frac{\partial S}{\partial \phi}. \
\qquad{A.4}
\end{equation*}
Integration yields:
\begin{equation*}
S(\xi',\phi) = k(\xi' \pm \phi) + \mathrm{constant}. \
\qquad{A.5}
\end{equation*}

Substituting back into (A.2), assuming $A$ approximately constant, the WKB solution becomes:
\begin{equation*}
\Psi(\xi',\phi) = A_0 e^{ik(\xi' \pm \phi)}. \
\qquad{A.6}
\end{equation*}

\subsection*{Appendix A.1. Hamiltonian Derivation in LQC Variables}

Starting with the gravitational Hamiltonian constraint:
\begin{equation*}
\mathcal{H}_G = -\frac{3c^2}{\gamma^2} \sqrt{|p|} + \frac{p_\phi^2}{2|p|^{3/2}} = 0,
\qquad{A.7}
\end{equation*}
In the above, $\gamma$ is the Immirzi parameter, in keeping with the discussion of LQC \cite{agullo2023loop}.  Multiplying the above equation by $2|p|^{3/2}$ gives:
\begin{equation*}
-6c^2 p^2 / (\gamma^2 ) + p_\phi^2 = 0.
\qquad{A.8}
\end{equation*}

Quantizing $c$ as:
\begin{equation*}
c \to i \frac{ \gamma}{3} \frac{\partial}{\partial p}, \
\qquad{A.9}
\end{equation*}
the Wheeler-DeWitt equation becomes:
\begin{equation*}
\frac{2}{3\gamma^2} p \frac{\partial}{\partial p}\left( p \frac{\partial \Psi}{\partial p}\right) + p_\phi^2 \Psi = 0. \
\qquad{A.10}
\end{equation*}

Introducing $p' = \ln p$, the derivatives transform as:
\begin{eqnarray*}
\frac{\partial}{\partial p} &=& \frac{1}{p} \frac{\partial}{\partial p'}, \nonumber \\
p \frac{\partial}{\partial p}p \frac{\partial}{\partial p} &=& \frac{\partial^2}{\partial p'^2}. \ \qquad{(A.11)}
\end{eqnarray*}

Thus, the Wheeler-DeWitt equation simplifies to:
\begin{equation*}
\frac{2}{3\gamma^2} \frac{\partial^2 \Psi}{\partial p'^2} + p_\phi^2 \Psi = 0. \
\qquad{A.12}
\end{equation*}
which can be rescaled to match the superspace D'Alembertian form:
\begin{equation*}
\frac{\partial^2 \Psi}{\partial p'^2} - \frac{\partial^2 \Psi}{\partial \phi^2} = 0.
\qquad{A.13}
\end{equation*}
\bibliographystyle{iopart-num}
\bibliography{ref}

\end{document}